# Pressurized rf cavities in ionizing beams

B. Freemire,[1,*] A. V. Tollestrup,[2] K. Yonehara,[2] M. Chung,[3] Y. Torun,[1]
R. P. Johnson,[4] G. Flanagan,[4] P. M. Hanlet,[1] M. G. Collura,[5] M. R. Jana,[6]
M. Leonova,[2] A. Moretti,[2] and T. Schwarz[7]

[1]*Illinois Institute of Technology, Chicago, Illinois 60616, USA*
[2]*Fermi National Accelerator Laboratory, Batavia, Illinois 60510, USA*
[3]*Uslan National Institute of Science and Technology, Uslan 689-798, Korea*
[4]*Muons, Inc., Batavia, Illinois 60510, USA*
[5]*University of California Santa Barbara, Santa Barbara, California 93106, USA*
[6]*Institute for Plasma Research, Bhat, Gandhinagar 3824 28, India*
[7]*University of Michigan, Ann Arbor, Michigan 48109, USA*

A muon collider or Higgs factory requires significant reduction of the six dimensional emittance of the beam prior to acceleration. One method to accomplish this involves building a cooling channel using high pressure gas filled radio frequency cavities. The performance of such a cavity when subjected to an intense particle beam must be investigated before this technology can be validated. To this end, a high pressure gas filled radio frequency (rf) test cell was built and placed in a 400 MeV beam line from the Fermilab linac to study the plasma evolution and its effect on the cavity. Hydrogen, deuterium, helium and nitrogen gases were studied. Additionally, sulfur hexafluoride and dry air were used as dopants to aid in the removal of plasma electrons. Measurements were made using a variety of beam intensities, gas pressures, dopant concentrations, and cavity rf electric fields, both with and without a 3 T external solenoidal magnetic field. Energy dissipation per electron-ion pair, electron-ion recombination rates, ion-ion recombination rates, and electron attachment times to $SF_6$ and $O_2$ were measured.

## I. INTRODUCTION

Muons are attractive particles to accelerate in high energy physics. They are 200 times as massive as the electron, and thus allow for the use of circular accelerators due to their relatively small energy loss through synchrotron radiation. Also, unlike protons, they are not composite particles, and so produce a much cleaner signal when they collide. Additionally, a single muon accelerator complex could provide both a high intensity, well-characterized neutrino factory and a multi-TeV muon collider.

As muons are unstable, they must be accelerated quickly. They are also created as tertiary particles, and thus require complex means of production and focusing before they may be used in an accelerator. Traditional methods of cooling beams of particles do not work within the lifetime of the muon, and ionization cooling appears to be the only viable alternative [1,2].

Ionization cooling works by passing a beam of particles through an energy absorbing material and replacing the lost longitudinal momentum with radio frequency (rf) cavities.

This technique requires the rf cavities to operate in strong external magnetic fields to provide a small beta function, which maximizes the cooling effect. It is important to note that angular spread of the beam must be larger than the angular spread due to scattering for cooling to be effective. To keep the cooling channel length short, a large cooling decrement is ideal, which dictates large voltage across the cavity in order to restore lost energy.

Past attempts to operate vacuum cavities in strong external magnetic fields have lead to problems with breakdown within the cavities [3–8]. It is believed that the magnetic field focuses field emission electrons from one wall of the cavity onto the opposing wall, causing an arc to form and short the cavity. Over time the energy deposited by such events fatigues the surface of the metal and causes irreparable damage. More recent work has provided some evidence that special cavity design and surface preparation techniques might alleviate breakdown in vacuum cavities, and that effort is progressing in parallel with the work presented here [9,10].

In order to mitigate breakdown, it was proposed that a rf cavity should be filled with a high pressure gas [11]. The gas acts as a buffer, reducing the mean free path of field emission electrons and preventing them from traversing the length of the cavity. Indeed, it has been shown that filling a rf cavity with a high pressure gas (a HPrf cavity) allows the cavity to operate without any performance reduction in external magnetic fields of 3 T [12,13].

[*]freeben@hawk.iit.edu

Hydrogen gas provides the best combination of radiation length and stopping power for ionization cooling, while also allowing for the operation of a rf cavity in strong magnetic fields. When a beam of muons passes through a HPrf cavity, it ionizes the gas. The resulting plasma is a source of free electrons and therefore for the HPrf cavity to be viable, the plasma electrons must not facilitate breakdown. Additionally, through collisions with gas molecules, the plasma will transfer energy from the cavity to the gas. This is known as plasma loading, and due to having a smaller mass, and therefore mobility, than ions, electrons are the main contributors.

This experiment was conducted with the intent to prove the feasibility of pressurized rf cavities for use in the cooling channel of a muon accelerator. A subset of the results for hydrogen and deuterium have been reported previously [14]. It is also hoped that this technology may be used in other applications. The test cell used in this experiment is not a prototype, however the physics results garnered may be used to extrapolate the performance of such a device to higher beam intensities and gas pressures.

To determine the total plasma loading expected at the higher beam intensity and gas pressure needed for a muon collider, the per-particle energy dissipation was measured (see Sec. IV). Past measurements of the mobilities and drift velocities of electrons and ions indicate that plasma loading may be an issue for intense muon beams, however no measurements have been made at the densities proposed.

Electrons will naturally recombine with hydrogen ions in a plasma, and the rate at which this happens depends on the plasma density and electric field. Electron-ion recombination results indicate that this process alone is not sufficient to support the beam intensities and time scales (tens of nanoseconds) currently under consideration for muon cooling (see Sec. V).

It is therefore necessary to dope with an electronegative gas to ensure the plasma electrons become attached to heavier molecules and thus significantly reduce the loading of the cavity (see Sec. VI). This process must occur within the nanosecond to sub-nanosecond time scale. Sulfur hexafluoride and oxygen were investigated, with oxygen being the ideal candidate due to $SF_6$ forming acids when reacting with hydrogen and having a high boiling point (it is desirable to operate the cooling channel at cryogenic temperatures—see Sec. VIII).

With sufficient concentrations of dopant, ions become the dominant contributor to plasma loading (see Sec. VII). Ions have been shown to recombine much slower than electrons, and it is clear that ion-ion recombination is not fast enough to significantly impact their population within the time frame of the bunch train. However between bunch trains (hundreds of microseconds to milliseconds) there is sufficient time to completely neutralize the gas.

## II. BEAM TEST OVERVIEW

By observing the electric field within the HPrf test cell when a beam passes through it, a great deal can be learned about the plasma physics processes that take place, through solving the plasma transport equations. The amount of energy each charged particle dissipates, the recombination rate of electrons with positively charged ions, the attachment time of electrons to electronegative molecules, and the recombination rate of ions must be measured in order to predict the total plasma loading in a real gas filled cooling channel. To this end, a variety of parent and dopant gases were studied, at different total pressures and concentrations, subject to different beam intensities [15]. The magnitude of the electric field within the test cell was also varied.

A dedicated beam line was constructed at the end of the Fermilab linac to service the MuCool Test Area. It provided a 400 MeV $H^-$ beam with a momentum spread of 0.005. The beam was bunched at 201 MHz, and traversed a ~100 m (drift) beam transport line before entering the experimental setup. The beam pulse length was variable, and was run between 7.5 and 9.5 $\mu$s. The total pulse intensity peaked at $\approx 2 \times 10^{12}$ protons.

The beam line ended approximately one meter upstream of the experimental setup. A 50.8 $\mu$m titanium window as the end of the beam line served to strip the electrons from $H^-$. The experimental apparatus was housed within the bore of a 3 T superconducting solenoid magnet. The beam passed through over one meter of air before entering the test cell within magnet bore.

A cross sectional view of the experimental setup is shown in Fig. 1. The beam hit a scintillating screen [16] mounted on the first of two stainless steel collimators, with through holes 20 and 4 mm in diameter. The collimator system allowed at most 1/5 of the total beam to reach the HPrf test cell, corresponding to an RMS beam size of $2.5 \times 4.0$ mm on the scintillating screen, and was used for beam quality and intensity selection. Two toroids with manganese zinc ferrite cores were mounted on the faces of the downstream collimator. The beam then passed through the HPrf test cell and ended in a stainless steel beam absorber. The entire apparatus was mounted on rails that were secured in place inside the bore of the superconducting solenoid magnet.

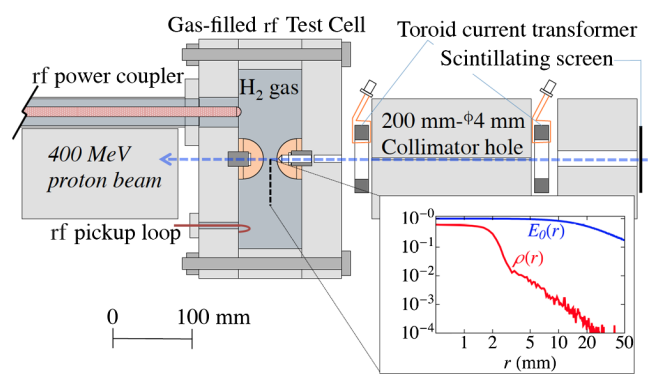

FIG. 1. The experimental setup housed within the bore of the superconducting solenoid. The beam enters from the right, passing through a scintillating screen and two collimators before impinging on the HPrf test cell. After traversing the test cell, the beam is stopped in a stainless steel cylinder. Toroids are mounted on both ends of the downstream collimator. The inset plot shows the radial distribution of the electric field amplitude and plasma density within the test cell.

TABLE I. HPrf test cell parameters.

| Parameter | Value | Units |
| --- | --- | --- |
| Resonant frequency (filled with $H_2$ gas) | 801.3–808.5 | MHz (103 – 20.4 atm) |
| Inductance | 26.08 | nH |
| Stored energy at 1 MV/m | 3.98 | mJ |
| Unloaded Q | 14,200 – 13,900 | (at 801–808 MHz) |
| Loaded Q | 6,900 – 6,400 | (at 801–808 MHz) |
| $R_0$ | 52.1 – 56.7 | Ω (at 801–808 MHz) |
| Cavity interior length | 8.13 | cm |
| Cavity interior diameter | 22.86 | cm |
| Electrode gap separation | 1.77 | cm |

The test cell was made of copper coated stainless steel and consisted of two end plates and a cylindrical body piece bolted together. A gasket between each plate and the body provided both the rf and pressure seal. Copper electrodes were used to enhance and localize the peak electric field on-axis. This structure also provided a high shunt impedance, increasing the sensitivity to plasma loading. The gap between the electrodes was 1.77 cm. The rf power coupler, gas line and all instrumentation were located on the downstream face of the test cell. The upstream face and electrode were counterbored to minimize the amount of material the beam had to interact with before entering the test cell (the thickness of material for each was 3.175 mm). Table I lists the parameters of the test cell.

The electric field distribution in the test cell was simulated using Superfish. The inset of Fig. 1 shows the radial distribution at the center of the cavity. Note that within the plasma column created by the beam, the electric field is nearly constant with radius, but varies by 30% over the length of the electrode gap.

Figure 2 illustrates the effect of plasma loading in a HPrf test cell. The rf envelope of five separate pulses are shown. In all but one case, the test cell was filled with hydrogen gas. The magenta data show a typical rf flat top with no beam. When the beam was turned on, significant plasma loading was observed. The addition of an electronegative gas greatly reduced the plasma loading.

### III. PLASMA FORMATION

Charged particles passing through a gas will interact with the gas through ionizing and dissociative ionizing collisions. In this experiment those particles are protons, however in this respect there is very little difference between the interactions of protons and muons. Single ionization is the dominant ionization process [17]:

$$p + H_2 \rightarrow p + H_2^+ + e^- \qquad (1)$$

with dissociative ionization occurring on the few percent level [17].

The ionization electrons can have enough energy to ionize hydrogen as well, increasing the number of electrons by ∼20%.

At high pressures of hydrogen background gas, the $H_2^+$ ions quickly interact to form $H_3^+$ (on the order of one picosecond) via [18]:

$$H_2^+ + H_2 \rightarrow H_3^+ + H \qquad (2)$$

Larger clusters of hydrogen can be formed through three-body collisions between the hydrogen ions and gas molecules [19]. The resulting hydrogen clusters can be dissociated through additional collisions.

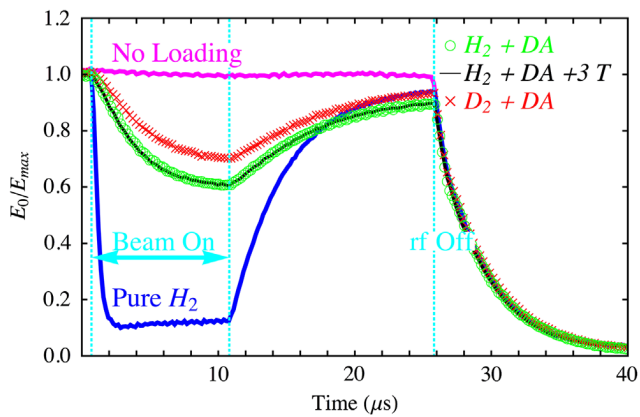

FIG. 2. Rf envelopes for various gas combinations. The beginning of the rf pulse has been omitted. The beam was sent through the test cell after the flat top electric field value had been reached. The magenta points are a rf pulse in which no beam was sent to the test cell. The blue points are a beam passing through the test cell filled with pure hydrogen. The green circles and black dashes represent hydrogen doped with dry air, both without and with a 3 T external magnetic field. The red crosses represent deuterium doped with dry air. The beam on and off, and rf off times are indicated by the dashed teal gridlines.

$$H_{n-2}^+ + 2H_2 \rightleftharpoons H_n^+ + H_2 \quad (n = 5, 7, 9, \ldots) \quad (3)$$

Eventually an equilibrium of the population of hydrogen ion clusters will be reached based on gas temperature and pressure. At the pressures of gas used in this experiment, the majority of ionic hydrogen is $H_5^+$ or larger [20].

The production rate of electron-ion pairs by the beam is calculated using the mass density of hydrogen gas, $\rho_m$, the average energy required to ionize a hydrogen molecule, $W_i$, the energy loss per unit length of 400 MeV protons in hydrogen $\frac{dE}{dx}$, and the propogation distance $h$:

$$\dot{N} = \dot{N}_b \times h \sum_k w_k \left( \rho_m \frac{dE/dx}{W_i} \right)_k \quad (4)$$

where $\dot{N}_b$ is the measured number of incident particles per unit time on the test cell, and the sum is taken over each $k$th gas species ($\sum_k w_k = 1$).

Energy gained by electrons from the rf field is transferred to the surrounding gas through collisions. Over the course of many collisions, the electrons will come into a thermal equilibrium above the gas temperature. The electron thermalization time is given by:

$$\tau_e = \frac{1}{\zeta_e \nu_e} \quad (5)$$

where $\nu_e$ is the collision frequency and $\zeta_e$ is the fractional energy loss per collision. For the case of electrons with energy below the ionization level, rotational and vibrational collisions dominate, and $\zeta_e \sim 10^{-3}$–$10^{-2}$ [21]. The energy loss for an elastic collision is smaller, $\zeta_e = 2m_e/(m_e + m_{H_2}) \approx 1/2000$. Assuming a Maxwellian distribution of electrons, the collision frequency over a pressure range of 20.4–103 atm (300–1520 psi) is 7.2–36.6 × $10^{12}$ s$^{-1}$ [22]. This gives a maximum thermalization time of 0.28 ns. Since the half period of 805 MHz is 0.62 ns, we will assume the electrons are always in thermal equilibrium with the surrounding gas. The electrons then drift with the applied rf electric field. It is assumed that diffusion is negligible over the time scales considered here.

## IV. PLASMA LOADING

The energy dissipated by a single particle, $dw$, can be estimated by integrating the dissipated power over a rf cycle:

$$dw = \int_{-T/2}^{T/2} P dt = 2q \int_0^{T/2} v_{\text{drift}} E dt$$

$$= 2q \int_0^{T/2} \mu[X_0 \sin \omega t](E_0 \sin \omega t)^2 dt \quad (6)$$

where $v_{\text{drift}}$ and $\mu$ are the drift velocity and the mobility of the particle (in which the dependence on $X$ has been shown), respectively, $E$ is the rf electric field, and $X_0$ is the electric field amplitude divided by the gas pressure.

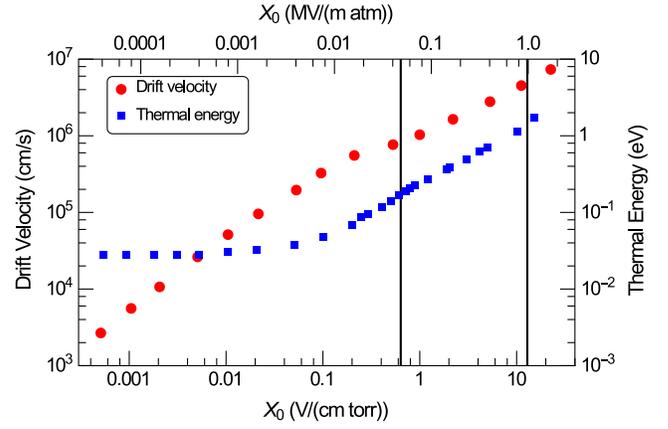

FIG. 3. Electron drift velocity and thermal energy in hydrogen gas vs $E/p$ [38]. The thermal energy has been derived from measurements of the electron diffusion constant and mobility, and the Einstein relation. The vertical lines represent the range of $X_0$ in the HPrf beam test.

The drift velocity of electrons in hydrogen gas has been well documented [23–30]. The mobility of electrons in hydrogen gas has also been well measured [31–37]. Figure 3 shows the drift velocity and thermal energy of electrons in hydrogen gas at 300 K vs $X_0$. $E/p$ is used because it specifies an electron's kinetic energy, i.e., if the electric field and the pressure are doubled, an electron will gain the same amount of energy between collisions. Below $X_0 \sim 0.05$ V/(cm torr) the electron's thermal energy is determined by the temperature of the gas (in this paper thermal energy is defined as $\frac{3}{2} k_b T$). The range of $X_0$ in the HPrf beam test is 0.636–11.6 V/(cm torr).

Equation (6) also applies to the ions present in the test cell, meaning that an estimate of the energy loss per ion can also be made. The mobilities of hydrogen clusters ($H_3^+$, $H_5^+$) and $O_2^-$ in hydrogen have been measured (Refs. [39–45] for hydrogen and [46] for oxygen). The mobilities of ions used in this work are given in Table II for reference.

The equivalent circuit for a beam and plasma loaded cavity is shown in Fig. 4. The generator (in our case a klystron) sends rf power down a matched transmission line, where it is inductively coupled to the cavity. The cavity can be modeled as a circuit with resistive, inductive, and capacitive components, with shunt impedance $R_c$. The gas is a source of energy loss and therefore acts as a resistive component. Due to the momentum spread of the beam and long drift before entering the experimental test cell, the beam arrives fairly continuously over the rf cycle, and therefore the average beam loading over multiple rf cycles is negligible.

TABLE II. Ion mobilities in hydrogen.

| Ion | Reduced mobility ($\frac{\text{cm}^2}{\text{V s}}$) |
|---|---|
| $H_3^+$ | 11.2 |
| $H_5^+$ | 9.6 |
| $O_2^-$ | 11.4 |

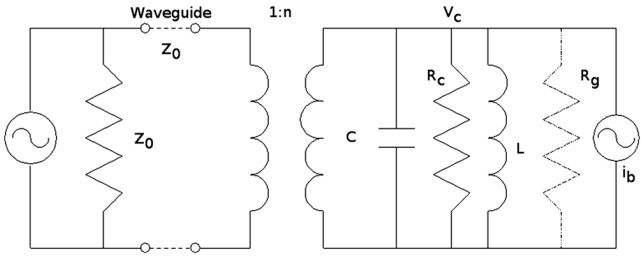

FIG. 4. The equivalent circuit of a matched cavity where the impedance, $Z_0$, of the waveguide and source reflect the resistance, $R_c$, of the cavity (after applying Thévinin's Theorem). The beam generated plasma within the cavity acts as a resistive component, $R_g$.

The power being delivered to the gas is given by:

$$P_g = \frac{\frac{1}{2}(V_0 - V)V}{\frac{1}{2}R_c} - \frac{d}{dt}\left(\frac{1}{2}CV^2\right) \quad (7)$$

where $V_0$ is the amplitude of the flat top cavity voltage, $V$ is the amplitude of the instantaneous voltage, and $C$ is the capacitance of the test cell. The voltage is found by using the average electric field across the accelerating gap. The first term is the power provided by the klystron, and the second term is the power provided by the test cell (coming from its stored energy).

The measured energy dissipated per electron-ion pair per rf cycle is found by dividing Eq. (7) by Eq. (4) and integrating over time. The predicted energy dissipation will be estimated using the particle's mobility or drift velocity and Eq. (6).

The measurement of $dw$ is dependent on the number of electrons produced within the test cell. Only very early times after the beam was turned on in pure gas were used, as recombination of electrons with hydrogen ions had not significantly changed the number of electrons present.

### A. Plasma loading results

Measurements of the energy dissipation per rf cycle per electron-ion pair were made for pure hydrogen, deuterium, nitrogen, and helium [47]. Figures 5, 6, and 7 show a summary of the results for each gas [48]. Lines representing the estimated energy dissipation due to electrons based on electron drift velocity measurements from the literature and Eq. (6) are also plotted. Systematic errors are on the order of 10% [49], while statistical errors are on the order of a few percent.

For a constant pressure, more energy is dissipated at higher $X_0$ (i.e., electric field), and for a constant $X_0$, more energy is dissipated at higher pressure. At low pressure the results match the predictions well. It can be seen that at high pressure, the measured values of the energy dissipation are less than the predicted values. Figures 8, 9, and 10 show the ratio of the measured to predicted value of $dw$ as a function of $X_0$ for hydrogen, deuterium, nitrogen, and helium. It has been previously observed that there is a pressure effect on the mobility and drift velocity of electrons in dense gases [30,32,33]. This is beneficial for the prospect of operating a muon cooling channel at higher gas pressure. It is worth pointing out that the trend observed here—a smaller mobility at higher pressure—is consistent with past experiments, however no prior data collected at room temperature in the pressure range explored here could be found.

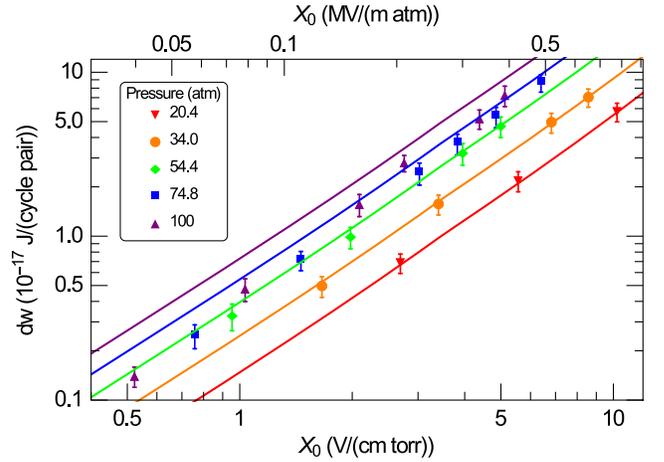

FIG. 5. Energy dissipation measurements for pure hydrogen. The lines represent predictions based on Eq. (6) and electron drift velocity measurements from Ref. [30].

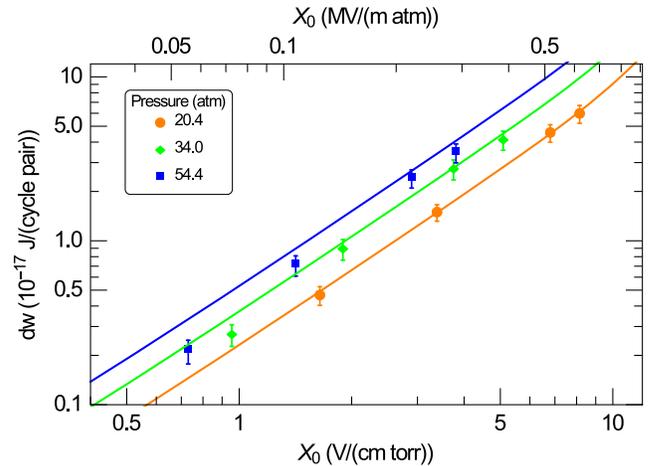

FIG. 6. Energy dissipation measurements for pure deuterium. The lines represent predictions based on Eq. (6) and electron drift velocity measurements from Ref. [50].

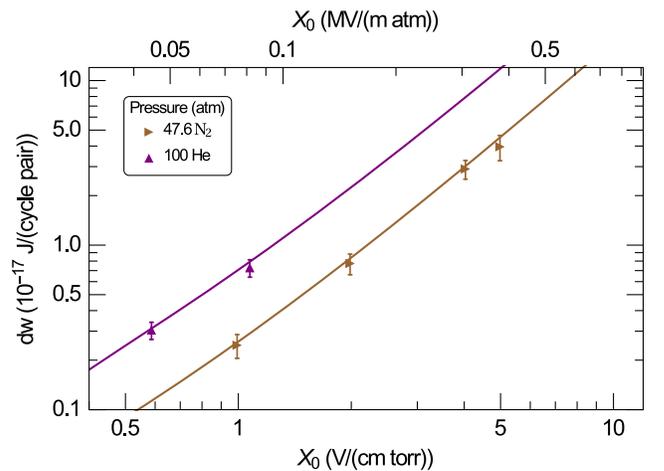

FIG. 7. Energy dissipation measurements for pure nitrogen and helium. The lines represent predictions based on Eq. (6) and electron drift velocity measurements from Refs. [30,51].

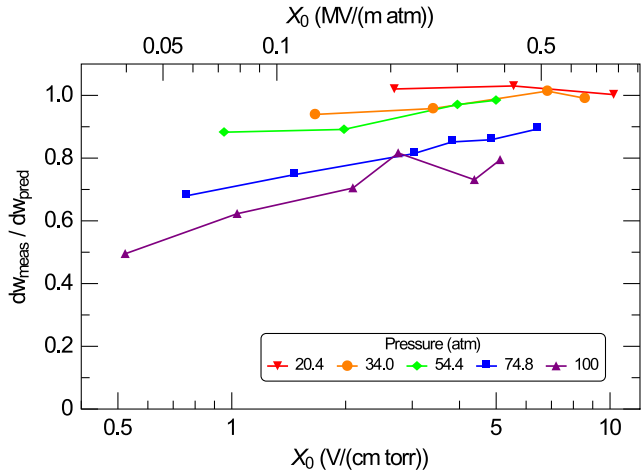

FIG. 8. Ratio of measured to predicted $dw$ vs $X_0$ for pure hydrogen [30].

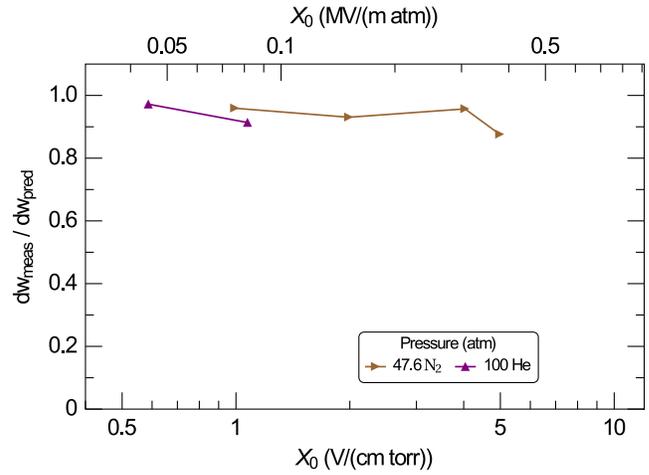

FIG. 10. Ratio of measured to predicted $dw$ vs $X_0$ for pure nitrogen and helium [30,51].

Energy dissipation measurements were also collected for doped gas combinations. Figure 11 shows the results for 20.4 atm hydrogen doped with varying concentrations of dry air. The lines on the plots represent two extreme cases:

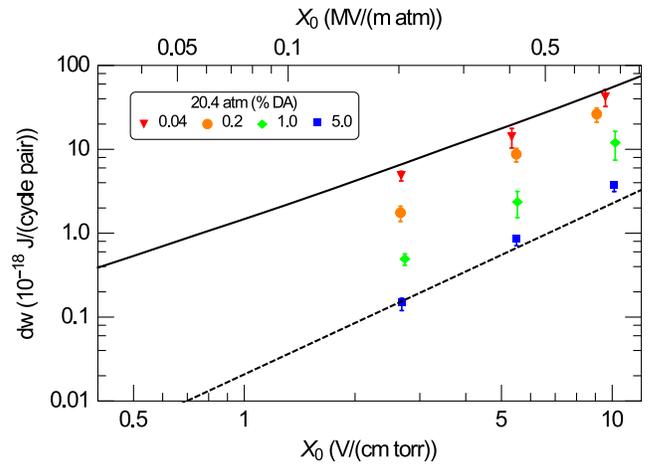

FIG. 11. Energy dissipation measurements for 20.4 atm hydrogen doped with varying concentrations of dry air. The black lines are predictions for only electrons (solid) and only ions (dashed) [30,46,52].

that in which the energy dissipation comes only from electrons, and that in which the energy dissipation comes only from ions ($H_5^+$ and $O_2^-$).

The data fall between the two extremes, indicating that the energy dissipation is coming from a combination of electrons and ions. At the smallest dopant concentration, the results are close to the prediction for only electrons, meaning a large number of electrons exist within the test cell. At the highest dopant concentration, the results are close to the prediction for only ions, meaning most of the electrons have become attached to an oxygen molecule, and ions largely determine the energy dissipation.

## V. ELECTRON-ION RECOMBINATION

Electrons must be removed as quickly as possible in order to minimize energy dissipation in the test cell. For the case of a pure gas, the only process through which this happens is recombination, which is frequently dissociative, with the final products being dependent on the initial hydrogen cluster, and gas temperature and density.

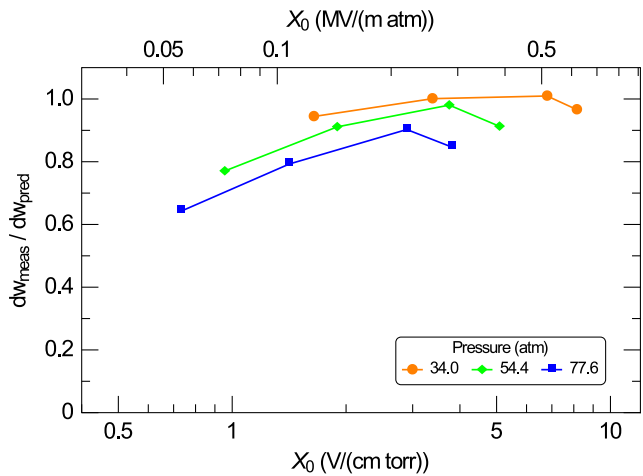

FIG. 9. Ratio of measured to predicted $dw$ vs $X_0$ for pure deuterium [50].

Rate equations for electrons and hydrogen ions are

$$\frac{dn_e}{dt} = \dot{n}_e - \sum_m \beta_m n_e n_{H_m^+} \qquad (8)$$

$$\frac{dn_{H_n^+}}{dt} = \dot{n}_{H_n^+} - \sum_m \beta_m n_e n_{H_m^+} \qquad (9)$$

where $n_\alpha$ is the number density of particle $\alpha$, $\dot{n}_\alpha$ is the production rate of particle $\alpha$, and $\beta_m$ is the recombination rate of electrons with $H_m^+$. If we assume a hydrogen ion is produced for every electron produced and there is no other means of removing electrons or ions, then Eqs. (8) and (9) reduce to:

$$\frac{dn}{dt} = \dot{n} - \beta n^2 \qquad (10)$$

where $\beta$ is the effective recombination rate for all hydrogen clusters. In Eqs. (8) through (10), the recombination rate is an average measurement over a rf cycle.

A measurement of the recombination rate was made at the minimum of the pure hydrogen rf curve in Fig. 2, as electrons were being produced at the same rate that they were recombining.

The recombination rates for both $H_3^+$ and $H_5^+$ have been measured extensively using a variety of methods ($H_3^+$ [53–57], $H_5^+$ [55–57]). There is approximately an order of magnitude difference in the recombination rates of $H_3^+$ and $H_5^+$. It has also been shown that as the electron temperature increases, the effective recombination rate decreases. However, the hydrogen gas densities in our experiment are many orders of magnitude larger than results previously published.

### A. Electron-ion recombination results

Measurements of the electron-ion recombination rate were made for pure hydrogen, deuterium, helium, and nitrogen [47]. Three beam intensities were recorded, and data were grouped into corresponding sets.

Figures 12, 13, and 14 show representative plots of the electron recombination rate with hydrogen, deuterium, nitrogen and helium ions, respectively, as a function of $X_0$ [48].

The exact species and population of ion could not be determined in this experiment. Consequently, the results presented here represent the effective recombination rate due to all ion species in the test cell.

As can be seen in Figs. 12 and 13, the recombination rate increases with increasing gas pressure and decreasing $X_0$.

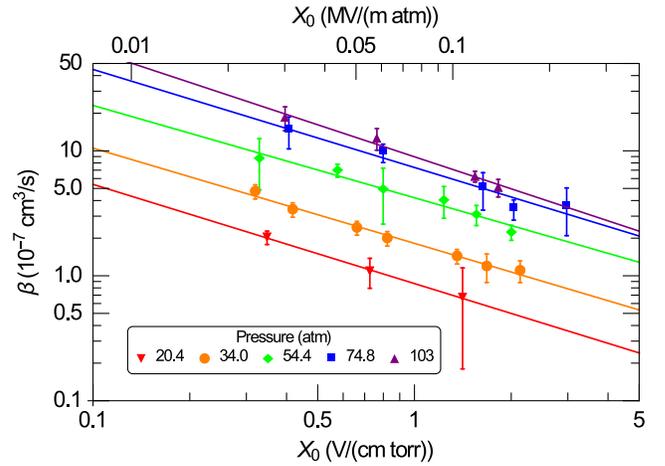

FIG. 12. Electron-ion recombination measurements vs $X_0$ for various pressures of hydrogen gas at the lowest beam intensity.

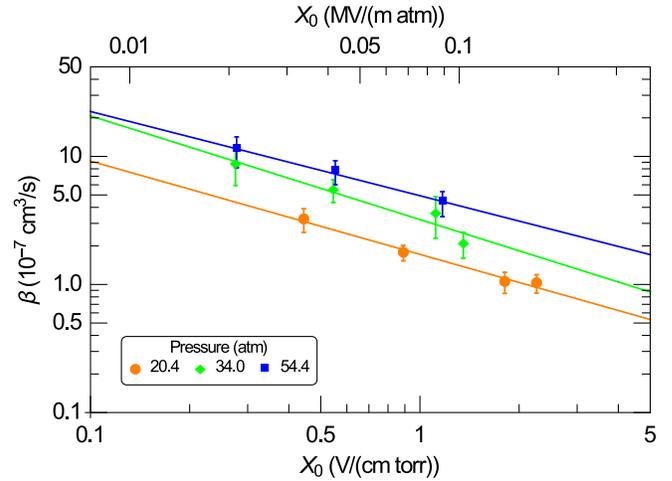

FIG. 13. Electron-ion recombination measurements vs $X_0$ for various pressures of deuterium gas at the lowest beam intensity.

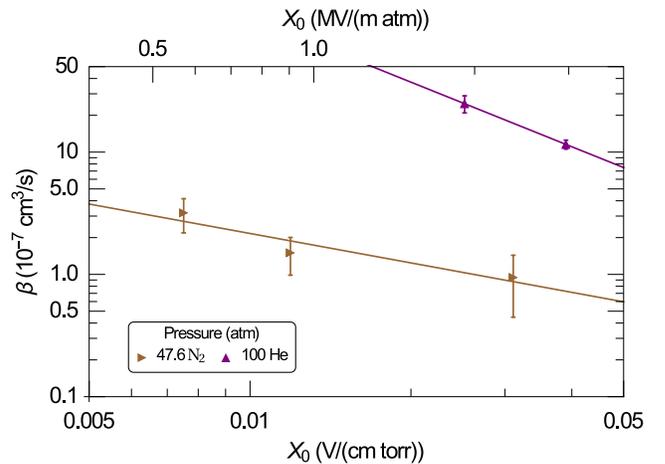

FIG. 14. Electron-ion recombination measurements vs $X_0$ for nitrogen and helium gases at the highest beam intensity.

This is not the case for all data, in particular the high beam intensity hydrogen data shows no clear pressure dependence. In recent hydrogen recombination experiments, saturation of the effective recombination rate was observed at a certain pressure, which varied with temperature [55]. This could also contribute to inconsistent pressure dependence in some data, or alternatively the large error is associated with taking measurements at low electric field [47].

Comparison of hydrogen and deuterium recombination rates show similar values. Recent work done by Novotný et al. involving a plasma of $D_3^+$ and $D_5^+$ ions is in good agreement with the results obtained here, and are consistent with hydrogen data as well [58].

Equilibrium constants for hydrogen and deuterium indicate the vast majority of ions present in this experiment are $H_5^+$ or $D_5^+$ and larger, for which there is no prior experimental recombination data in the pressure range reported here. Nonetheless, the data presented here are consistent with trends from past experiments: increasing rates with gas density and decreasing electron temperature, and values $10^{-7}$–$10^{-6}$ cm$^3$/s.

## VI. ELECTRON ATTACHMENT

Measurements of the recombination rate of electrons in hydrogen reported here indicate that recombination is not fast enough to sufficiently remove electrons during the nanosecond bunch spacing of a muon accelerator. An electronegative gas must therefore be used to minimize the plasma loading due to electrons by effectively decreasing their mobility through attachment to a molecule. It can be seen in Fig. 2 that the addition of oxygen (dry air) significantly reduces the plasma loading.

Electron attachment to oxygen is a three-body process and involves two steps [59]. In the first, an oxygen molecule captures an electron, resulting in an excited state:

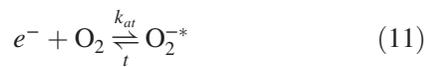
$$e^- + O_2 \underset{t}{\overset{k_{at}}{\rightleftharpoons}} O_2^{-*} \qquad (11)$$

where $k_{at}$ is the attachment rate of $O_2^{-*}$ formation, and $t$ is the lifetime of $O_2^{-*}$ before it decays into the initial particles. One of two things can take place at this point. Either the oxygen can be deexcited by a collision with another gas molecule ($M$), or ionized:

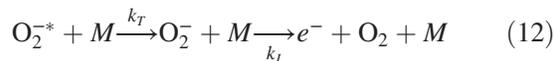
$$O_2^{-*} + M \xrightarrow{k_T} O_2^- + M \underset{k_I}{\longrightarrow} e^- + O_2 + M \qquad (12)$$

Here, $k_T$ is the rate of deexcitation, and $k_I$ is the rate of ionization.

Finally, the attachment coefficient for the three-body process [Eq. (11)) and Eq. (12)] depends on $k_{at}$ multiplied by the probability that $O_2^{-*}$ will deexcite:

$$k_m = \frac{k_{at}k_T}{t^{-1} + (k_T + k_I)n_M} \qquad (13)$$

where $n_M$ is the density of the third body. The collision frequency is sufficiently high for the gas pressures explored here that the excited state of oxygen is extremely likely to experience a collision within the length of time required to decay into the initial particles.

The rate equations for electrons, hydrogen ions, and oxygen ions are

$$\frac{dn_e}{dt} = \dot{n}_e - \sum_l \beta_l n_e n_{H_l^+} - \sum_m k_m n_e n_m n_{O_2^-} \qquad (14)$$

$$\frac{dn_{H_n^+}}{dt} = \dot{n}_{H_n^+} - \sum_l \beta_l n_e n_{H_l^+} - \sum_l \eta_l n_{H_l^+} n_{O_2^-} \qquad (15)$$

$$\frac{dn_{O_2^-}}{dt} = \sum_m k_m n_e n_m n_{O_2^-} - \sum_l \eta_l n_{H_l^+} n_{O_2^-} \qquad (16)$$

where the sum over $l$ is for each cluster of hydrogen, and the sum over $m$ is for each species of gas molecule. The effective lifetime of an electron is given by

$$\tau = \frac{1}{\sum_m k_m n_m n_{O_2^-}} \qquad (17)$$

where $k_m$ is the same as in Eq. (13).

Most past measurements have been made in pure oxygen [60–69], or in oxygen-nitrogen mixtures [60–63,65–67]. Only a few sources report the attachment coefficient in an oxygen-hydrogen mixture [61,70]. These data were collected at a single gas pressure and electron temperature. However, the pressure dependence have been measured for pure oxygen, and oxygen-nitrogen and oxygen-helium mixtures. In those cases, the rate of attachment increases with gas pressure. Additionally, the attachment coefficient has been shown to decrease with electron temperature ($X_0$) [63].

Table III shows selected three-body attachment coefficients for thermal electrons in various parent gases at 300 K.

Using Eq. (17) and the values for the attachment coefficients of hydrogen, nitrogen, and oxygen from Table III, the attachment time for electrons to oxygen in dry air doped hydrogen at room temperature can be calculated for thermal electrons. Table IV lists the results for 20.4 and 100 atm hydrogen doped with dry air concentration from 0.001 to 5%.

TABLE III. Various three-body attachment coefficient measurements of electrons to oxygen in various gases at 300 K [61,62,66,68–70].

| Gas | Attach. coeff. ($10^{-31}$ $\frac{cm^6}{s}$) |
|---|---|
| $H_2$ | 2.0, 4.8 |
| $O_2$ | 20, 21.2, 25, 28 |
| $N_2$ | 1.0, 1.1, 1.6 |

TABLE IV. Calculated electron attachment times to oxygen using Eq. (17) and values of the attachment coefficient from Table III. Listed are the attachment times for dry air doped hydrogen at two gas pressures and various concentrations.

| Dry air concentration (%) | $\tau$ (ns) | |
| --- | --- | --- |
| | 20.4 atm | 100 atm |
| 0.001 | 4130 | 172 |
| 0.002 | 2070 | 86.0 |
| 0.04 | 103 | 4.30 |
| 0.2 | 20.6 | 0.860 |
| 1 | 4.12 | 0.172 |
| 5 | 0.814 | 0.0339 |

## A. Electron attachment results

Results for the characteristic time for an electron to become attached to an electronegative molecule were obtained for: hydrogen doped with dry air, sulfur hexafluoride, and nitrogen, as well as deuterium, helium, and nitrogen doped with dry air, at varying concentrations and pressures [47]. Two beam intensities were used during data taking. Only data from the highest beam intensity will be presented here [48].

Figures 15 and 16 show measurements of the attachment time as a function of $X_0$ for 20.4, and 98.6 and 100 atm, respectively, of hydrogen doped with various concentrations of dry air. The resolution on the attachment time is 1 ns [47]. As a result, for the cases of high pressure and dopant concentration, this data set can only place an upper limit on the value of $\tau$. Such data points are represented by open symbols, do not have error bars, and have not been included in the fits. The solid lines in the following plots are fits to the data, while the dashed horizontal lines are predictions of the attachment time for a given dry air concentration.

Figure 17 shows an extrapolation of how $\tau$ varies with hydrogen gas pressure at a fixed $X_0$ for 0.04, 0.2, and 1% dry air. The points on this plot are obtained through fits of $\tau$ as a function of $X_0$ for 20.4, 74.8, and 100 atm. Additionally, for 100 atm, a fit of $\tau$ as a function of dry air concentration was used.

Figure 18 compares how $\tau$ varies with $X_0$ for hydrogen and deuterium at 20.4 atm and 1% dry air. When the difference in electron temperature as determined by $X_0$ in hydrogen versus deuterium is accounted for, the values for $\tau$ are in very good agreement.

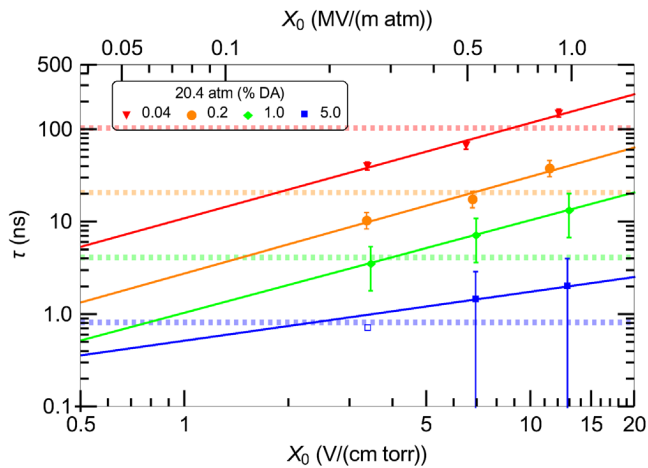

FIG. 15. Electron attachment time as a function of $X_0$ at 20.4 atm for various dry air concentrations. The estimated error is 10% of the measurement value at 0.04% DA, 20% at 0.2% DA, 50% at 1% DA, and 100% at 5% DA.

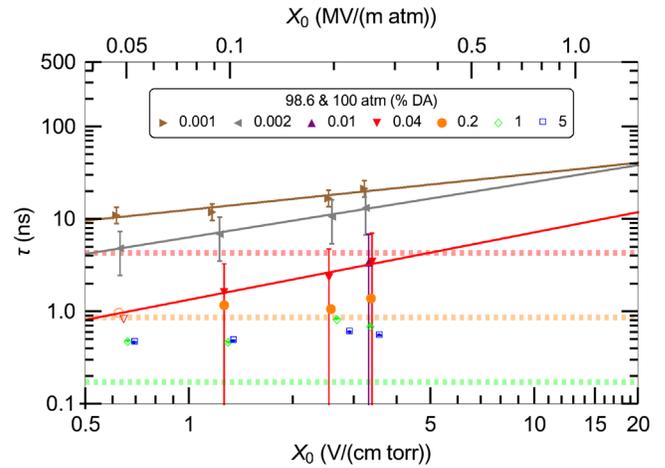

FIG. 16. Electron attachment time as a function of $X_0$ at 98.6 and 100 atm for various dry air concentrations. The estimated error is 20% of the measurement value at 0.001% DA, 50% at 0.002% DA, 100% at 0.01% DA, and 100% at 0.04% DA. The data points for larger concentrations can only be considered upper limits on the attachment time.

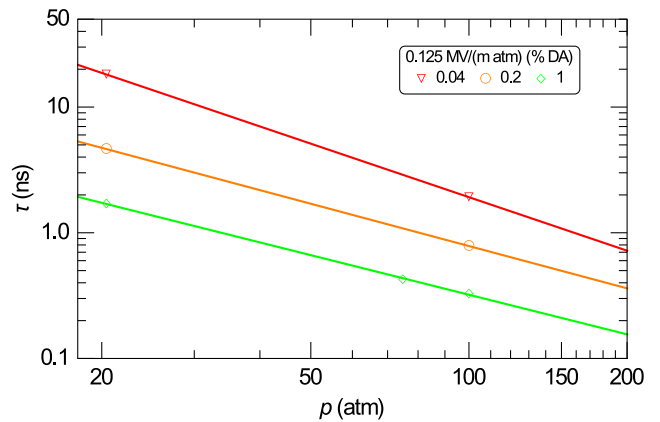

FIG. 17. Electron attachment time as a function of $p$ at $X_0 = (20 \text{ MV/m})/(160 \text{ atm})$ for various dry air concentrations.

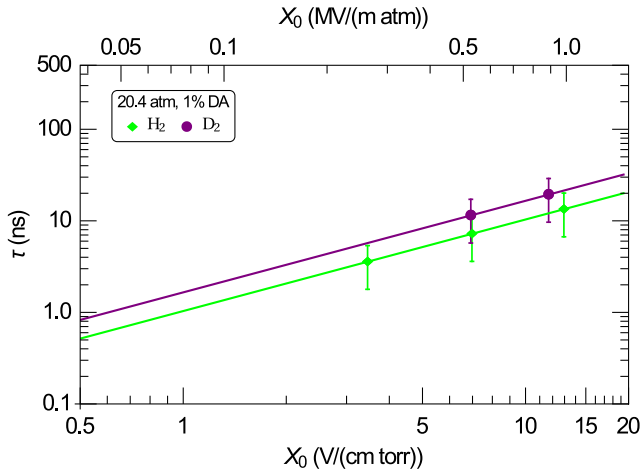

FIG. 18. Electron attachment time as a function of $X_0$ at 20.4 atm and 1% dry air for deuterium and hydrogen. The estimated error is 50% of the measurement value.

Figure 19 compares how $\tau$ varies with dry air concentration for hydrogen and nitrogen at 2.76 V/(cm torr). Data were not recorded at the same pressure for each gas, and so 20.4 atm is shown for hydrogen and 47.6 atm is shown for nitrogen.

Figure 20 shows how $\tau$ varies with $X_0$ for 1% dry air doped helium at 100 atm.

Sulfur hexafluoride is extremely effective at capturing electrons. With the exception of two data points, all measurements of the attachment time of electrons to $SF_6$ were smaller than the 1 ns precision. These two data points were 2.00 and 1.99 ns, measured for $X_0 \approx 1$ MV/(m atm), $p = 20.4$ atm, and concentrations of 0.00004% and 0.0002%, respectively.

Comparing the calculated values of $\tau$ from the literature given in Table IV with the measured values for 20.4 and 100 atm dry air doped hydrogen, there is good agreement for concentrations greater than 0.002%.

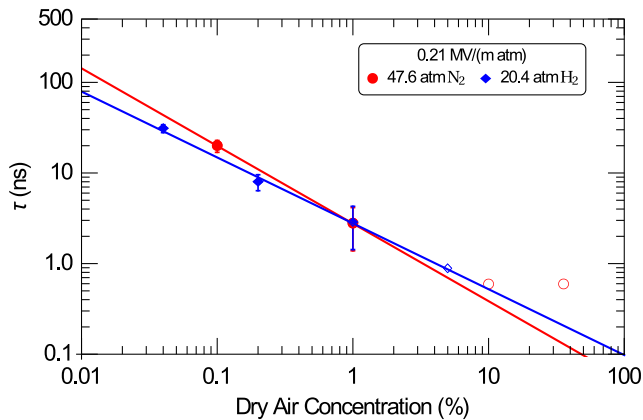

FIG. 19. Electron attachment time as a function of dry air concentration at $X_0 = 2.76$ V/(cm torr) for nitrogen gas at 47.6 atm and hydrogen gas at 20.4 atm.

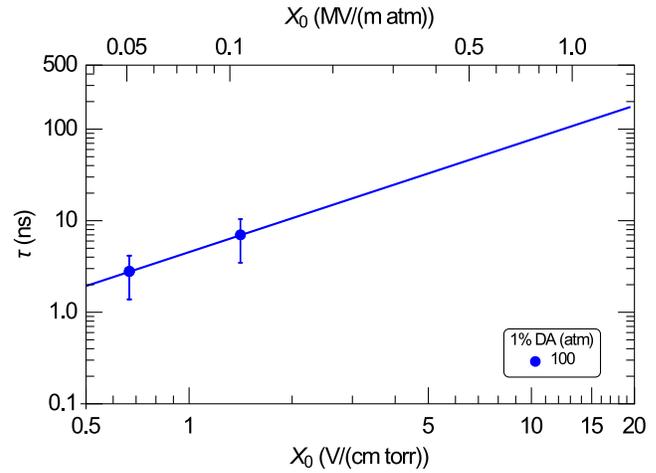

FIG. 20. Electron attachment time as a function of $X_0$ at 100 atm for 1% dry air doped helium.

Electrons become attached to oxygen much more quickly in hydrogen as compared to nitrogen or helium.

In all cases the following trends are true: the attachment time increases with electron temperature ($X_0$), decreases with dopant concentration, and decreases with gas pressure. At the large gas densities in this experiment, the three body attachment process looks very much like a two body process, as the excited state of oxygen almost immediately has a collision with another molecule. This is supported by the almost linear dependence of the attachment time to gas pressure. It is known that the attachment coefficient varies with electron temperature and indeed, the attachment time is nearly linear with $X_0$ for the dry air hydrogen and deuterium data.

## VII. ION-ION RECOMBINATION

Ions also contribute to plasma loading. The rate at which they neutralize is much slower than electron recombination, and will give an indication as to the long term evolution of the plasma.

For pressures above 1 atm, ion-ion recombination can be described by the Langevin model [71]. It is a three-body process similar to that of electron attachment

TABLE V. Selected ion-ion recombination rates [72–77].

| Reaction | Rate ($10^{-8}\frac{cm^3}{s}$) | Gas density or temperature |
|---|---|---|
| $O_2^- + O_4^+ + O_2 \rightarrow 4O_2$ | 420 | $2.7 \times 10^{19}$ cm$^{-3}$ |
| $O_2^- + O_4^+ + O_2 \rightarrow O_6 + O_2$ | 220 | $2.7 \times 10^{19}$ cm$^{-3}$ |
| $O_2^- + O_4^+ + O_2 \rightarrow O_6 + O_2$ | 30 | $5.4 \times 10^{20}$ cm$^{-3}$ |
| $O_2^- + O_2^+ \rightarrow O_2 + O_2^*$ | 14 | 200 K |
| $O_2^- + O_2^+ \rightarrow O_2 + O_2^*$ | 8.92 | 500 K |
| $H^- + H^+ \rightarrow$ | 3.9 | Thermal |
| $O_2^- + N_2^+ \rightarrow$ | 16 | Thermal |
| $H^- + H^+ \rightarrow$ | 40 | 300 K |

$$A^- + B^+ + M \rightarrow [AB] + M \qquad (18)$$

where the brackets indicate dissociation is possible. Very little data exists at high gas density, however Bates and Flannery [72] were able to successfully modify the Langevin-Harper formula

$$k_{ii} = \frac{4\pi e}{\epsilon_0}(\mu_+ + \mu_-) \qquad (19)$$

which gives the ion-ion recombination rate based on the mobilities of the positive and negative ions, to closely match high density oxygen ion-ion recombination data. This was accomplished by correctly accounting for the mean free paths of each ion species, and modifying the classical ion mobility to match measurements made at higher density. The result is that ion-ion recombination rates peak around 1 atm, and falls off at higher pressures.

No data on the ion-ion recombination rates of $H_5^+$ or $H_7^+$ with $O_2^-$ could be found. A variety of other ion rates are available, and are listed in Table V [72–77].

### A. Ion-ion recombination results

The ion-ion recombination rate, $\eta$, is derived from Eqs. (15) and (16) [47]. Representative plots are shown below [48]. Note that due to the analysis method used to obtain the electron attachment time and ion-ion recombination rate, the two measurements are coupled. Greater emphasis was placed on the accuracy of the electron attachment time, therefore the ion-ion recombination rate has a larger error. The errors are typically in the range of 50–100% of the measured value. Data collected in which it is believed that the majority of electrons are quickly captured yield more accurate results.

Figure 21 shows the recombination rate of $H_n^+$ and $O_2^-$ as a function of $X_0$ for 100 atm hydrogen doped with varying amounts of dry air. The lines on the plots are fits to the data.

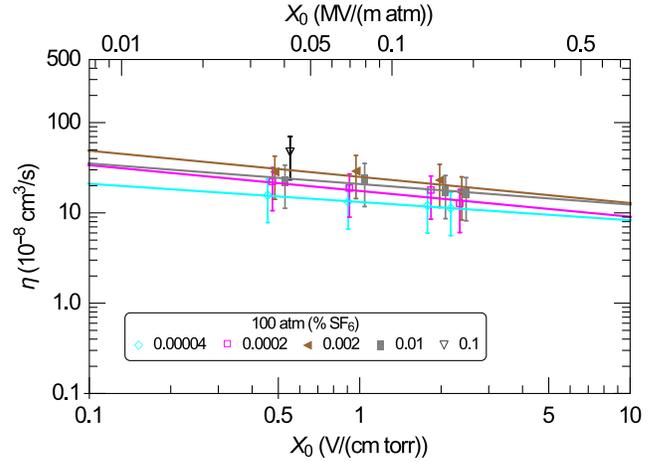

FIG. 22. Ion-ion recombination rate as a function of $X_0$ at 100 atm $SF_6$ doped hydrogen.

Figure 22 shows the recombination rate of $H_n^+$ with $SF_6^-$ as a function of $X_0$ for varying concentrations of $SF_6$ doped hydrogen at 100 atm.

Figure 23 shows the recombination rate of $D_n^+$ with $O_2^-$ as a function of $X_0$ for varying pressures of 1% deuterium doped hydrogen.

The ion-ion recombination rates are roughly independent of gas pressure and dopant concentration for this range of pressure and $X_0$. There may be a slight inverse relationship with $X_0$. The most pure measurement made (i.e. fewest electrons present) is in the largest concentration $SF_6$ doped hydrogen data, for which the recombination rate is a few times $10^{-7}$ cm$^3$/s. In the dry air doped hydrogen data, it can be seen that as electrons are removed (i.e. increasing concentrations of dry air), the recombination rate settles down to $1-2 \times 10^{-8}$ cm$^3$/s. It is interesting to note that the values for the recombination rate for 1% dry air doped deuterium at 100 atm are roughly 10 times larger than the same pressure and concentration for dry air doped hydrogen.

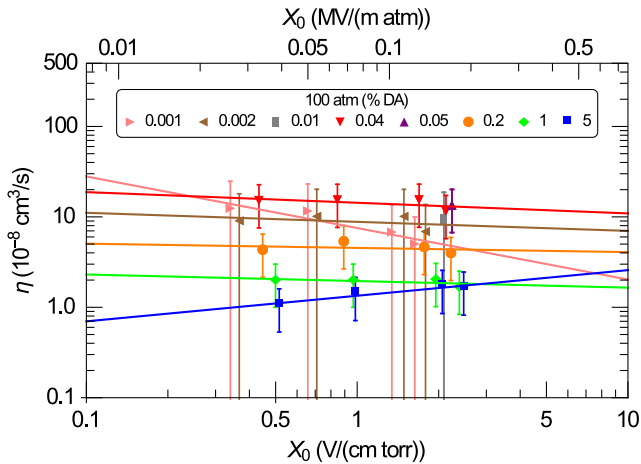

FIG. 21. Ion-ion recombination rate as a function of $X_0$ at 100 atm dry air doped hydrogen.

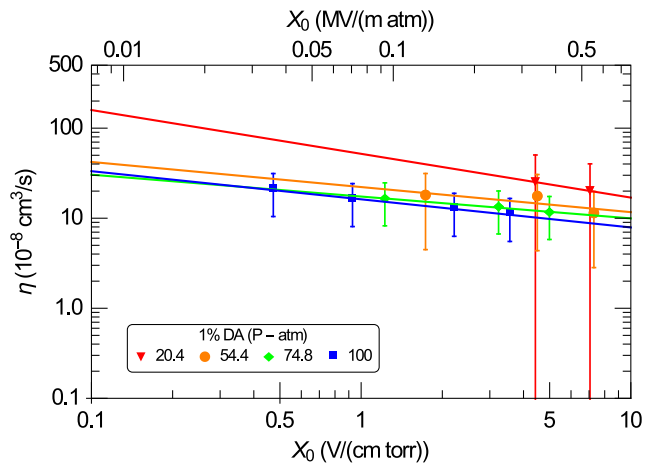

FIG. 23. Ion-ion recombination rate as a function of $X_0$ at 20.4 atm for 1% dry air doped deuterium.

## VIII. APPLICATION TO A MUON COOLING CHANNEL

The application of these results to one cooling scheme will be discussed here. Ionization cooling intrinsically only cools transversely. In order to achieve six dimensional cooling, an emittance exchange mechanism must be employed. The helical cooling channel (HCC) accomplishes this by using a magnetic channel comprised of magnetic solenoidal, helical transverse dipole and quadrupole fields, with hydrogen gas filled rf cavities placed along the beam's helical orbit [78]. This forces higher momentum particles to traverse more cooling medium, providing the energy loss, reacceleration, transverse cooling, and emittance exchange continuously along the length of the channel.

Table VI lists the design parameters for the beam and HCC. The results reported in this paper have been applied to this cooling channel scheme [79]. Here, it is important to note that one beam pulse is comprised of 21 bunches.

The expected plasma loading as a function of incident muon bunch intensity has been calculated based on the parameters in Table VI and the energy dissipation, recombination, and attachment results presented in this paper. Figure 24 shows the percent of the stored energy of each frequency cavity that is expected to be dissipated by plasma loading as a function of bunch intensity.

Based on the beam and cooling channel parameters listed above, and extrapolations of the plasma physics results presented here, the maximum bunch intensity should be below $4 \times 10^{12}$ muons for the 325 MHz section, and $10^{12}$ muons for the 650 MHz section of the HCC ($8.2 \times 10^{13}$ and $2.1 \times 10^{13}$ total muons per pulse, respectively). Operating the cavities at colder temperatures may alleviate some plasma loading, as colder electrons have higher recombination and attachment rates.

Similar studies have been carried out for another cooling channel scheme, the hybrid rectilinear channel, as well as the front end section of the cooling channel [80]. The results reported here will guide decisions to be made on beam intensity, gas pressure, and electric field for each cooling channel design.

TABLE VI. Helical cooling channel design and beam parameters [79].

| Parameter | Unit | Value |
|---|---|---|
| Rf frequency | MHz | 325, 650 |
| Gas species |  | Hydrogen |
| Gas pressure | atm | 180 |
| Oxygen concentration | % | 0.2 |
| Peak electric field | MV/m | 20 |
| External magnetic field | T | 4–14 |
| Number of bunches |  | 21 |
| Bunch frequency | MHz | 325 |
| Injection phase | degrees | 160 |

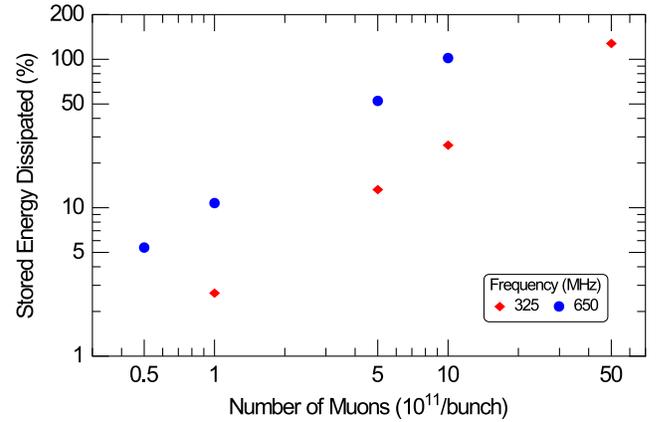

FIG. 24. Percent of the total stored energy for 325 and 650 MHz cavities dissipated due to plasma loading in the HCC. One beam pulse is comprised of 21 bunches.

## IX. CONCLUSION

The conditions inside the HPrf test cell when the data reported here were taken are unique, and represent a region previously uninvestigated for this combination of gas pressure, plasma density, and electric field.

There are four processes of interest when evaluating an HPrf cavity: energy dissipation due to electrons and ions; election-ion recombination; electron attachment to an electronegative molecule; and ion-ion recombination. The expected degradation in accelerating gradient in such a cavity due to beam-induced plasma is the main concern. How the plasma evolves through the electron-ion, electron-electronegative gas, and ion-ion interactions dictates the level of plasma loading.

The energy dissipation of electrons and ions in hydrogen, deuterium, helium, and nitrogen gas has been measured as functions of gas pressure and electric field. Predictions of the energy dissipation per electron-ion pair made using past measurements of electron drift velocity and ion mobility match the data collected here well for small pressures. At larger pressures a "saturation" of energy dissipation was observed. Previously reported drift velocity and mobility dependence on gas pressure support such an observation.

Electron recombination rates to clusters of hydrogen, deuterium, helium and nitrogen have been measured. While the exact cluster population is unknown, the measured rates are consistent with previous measurements of $H_3^+$, $H_5^+$, $D_3^+$, and $D_5^+$ mixtures. The effective recombination rates as a function of gas species and pressure, and electric field were reported.

The attachment time of an electron to oxygen or sulfur hexafluoride in parent gases of hydrogen, deuterium, helium, or nitrogen was measured. Past experimental data on the attachment coefficient for hydrogen and deuterium taken at low pressure is in good agreement with the data collected here. The attachment time as a function of gas species and pressure, dopant species and concentration, and electric field were reported.

Ion-ion recombination rates for hydrogen, deuterium, helium, and nitrogen ions with oxygen or sulfur hexafluoride ions have been measured. While the exact cluster size of the positive ion is unknown, the species of ions appears to be the largest factor in determining the rate. Unfortunately no prior experimental data exists to corroborate these findings.

An estimation of the plasma loading in one proposed muon cooling channel schemes has been reported based on the data presented in this paper, and others have been estimated elsewhere [79,80]. The successful application of high pressure gas filled rf cavities in muon cooling channels for bright muon sources relies on extrapolation of cavity performance to an environment of larger beam intensity and higher gas and plasma density. To this end a simulation package is being developed to apply the physics results garnered here to such regimes [81,82]. Initial indications are that gas filled rf cavity technology could be successfully applied to a cooling channel for a bright muon source.

## ACKNOWLEDGMENTS

The authors would like to thank the Muon Accelerator Program for making this work possible, Muons, Inc. for supplying the high pressure test cell for this experiment, the Linac crew at Fermilab for providing beam line support, and all the MTA personnel involved. Special thanks goes to Rainer Johnsen of the University of Pittsburgh, who contributed invaluable insight and with whom we had many useful discussions. Fermilab is operated by Fermi Research Alliance, LLC under Contract No. DE-AC02-07CH11359 with the U.S. Department of Energy.